\documentclass[twocolumn]{revtex4-1}
\usepackage{graphicx}
\usepackage{bm}
\usepackage{graphicx}
\usepackage{amsmath}
\usepackage{mathrsfs}
\usepackage{ulem}
\usepackage{color}
\usepackage{booktabs}
\usepackage{multirow}
\usepackage{diagbox}
\usepackage[colorlinks, linkcolor=red,anchorcolor=green,citecolor=blue]{hyperref}

\graphicspath{{./figs/}}
\begin{document}   

\title{Magnetic field induced hair structure in charmonium gluon-dissociation}
\author{Jin Hu$^1$}
\author{Shuzhe Shi$^{2,3}$}
\author{Zhe Xu$^1$}
\author{Jiaxing Zhao$^1$}
\author{Pengfei Zhuang$^1$}
\affiliation{$^1$Physics Department, Tsinghua University, Beijing 100084, China\\
             $^2$Department of Physics, McGill University, Montre\'al, QC H3A 2T8, Canada\\
             $^3$Center for Nuclear Theory, Department of Physics and Astronomy, Stony Brook University, Stony Brook, New York, 11784, USA}

\begin{abstract}
We study electromagnetic field effect on charmonium gluon-dissociation in quark-gluon plasma. With the effective Hamiltonian derived from QCD multipole expansion under an external electromagnetic field, we first solve the two-body Schr\"odinger equation for a pair of charm quarks with mean field potentials for color and electromagnetic interactions and obtain the charmonium binding energies and wave functions, and then calculate the gluon-dissociation cross-section and decay width by taking the color electric and magnetic dipole interactions as perturbations above the mean field and employing Fermi's Golden Rule. Considering the charmonium deformation in magnetic field, the discrete Landau energy levels make the dissociation cross-section grow hair, and the electric dipole channel is significantly changed, especially for the $P-$wave states $\chi_{c0}$ and $\chi_{c\pm}$. From our numerical calculation, the magnetic field strength $eB=5~m_\pi^2$ already changes the gluon dissociation strongly, which may indicate measurable effects in high-energy nuclear collisions.           
\end{abstract}
\maketitle

%==================================
\section{Introduction}
\label{sec1}
It is widely accepted that the strongest electromagnetic field in nature can be created in non-central relativistic heavy-ion collisions~\cite{Skokov:2009qp,Voronyuk:2011jd,WTDeng:2012prc,Tuchin:2013ie}. In Au-Au collisions at Relativistic Heavy Ion Collider (RHIC) the peak value of the magnetic field is around $eB\sim 5~m_{\pi}^2$, and in Pb-Pb collisions at Large Hadron Collider (LHC) the value even reaches $eB\sim 70~m_{\pi}^2$~\cite{WTDeng:2012prc}, where $m_\pi$ is the pion mass in vacuum. While such strong electromagnetic field can bring us many fantastic topics in quantum chromodynamics (QCD) physics, such as chiral magnetic effect~\cite{Kharzeev:2008npa,Fukushima:2008xe} and inverse magnetic catalysis~\cite{Shovkovy:2012zn,Bruckmann:2013oba}, the initially produced field decays very fast and survives only in the very beginning of the collisions, although the attenuation is delayed slightly as quark-gluon plasma (QGP) appears afterward~\cite{Tuchin:2013ie,Gursoy:2014aka,Yan:2021zjc,Chen:2021nxs,Wang:2021oqq}. 

Heavy quarks are probably an ideal probe of the short-lived electromagnetic field due to the fact that they are produced at the very early stage of heavy-ion collisions too. The difference in the directed flow between $D^0$ and $\bar D^0$ may come from the electromagnetic field~\cite{Adam:2019wnk,Acharya:2019ijj,Das:2016cwd}, and the quarkonium static properties such as the mass and shape are changed sizeably in the field~\cite{Marasinghe:2011bt,Alford:2013jva,Machado:2013rta,Cho:2014exa,Guo:2015nsa,Bonati:2015dka,Bonati:2017uvz,Yoshida:2016xgm,Zhao:2020jqu,Mishra:2020kts,Chen:2020xsr,Iwasaki:2021nrz}. The field affects also the quarkonium dissociation in hot medium~\cite{Singh:2017nfa,Hasan:2018kvx,Hasan:2017fmf,Hasan:2020iwa}. Different from the color screening picture~\cite{Karsch:1987pv} based on calculations at mean field level, the dissociation processes which originate from the scattering between quarkonia and thermal partons might be realistic dynamics for quarkonium suppression in high energy nuclear collisions. There are two kinds of dissociation processes, one is gluon dissociation ($g+\Psi \to Q+\bar Q$), the other is inelastic parton scattering ($p+\Psi \to Q+\bar Q +p$), where $g$ and $p$ represent gluons and partons. The former is dominant in the temperature region where the Debye mass is much smaller than the binding energy, and the latter is essential when the quarkonium becomes a loosely bound state~\cite{Brambilla:2013dpa,Brambilla:2011sg}. When the external electromagnetic field is turned on, the Landau-damping leads to an increasing decay width in the inelastic scattering processes~\cite{Singh:2017nfa,Hasan:2018kvx,Hasan:2017fmf,Hasan:2020iwa}.

The gluon dissociation describes the process of a color-singlet state converting to a color-octet state by absorbing a gluon~\cite{Brambilla:2011sg}. The cross section in vacuum neglecting the color-octet interaction in the final state was calculated firstly by Bhanot and Peskin via the operator-product-expansion (OPE) method~\cite{Peskin:1979va,Bhanot:1979vb}. Peskin's perturbative analysis can be represented by a gauge-invariant effective action from which one can get a non-relativistic Hamiltonian for heavy quark systems via QCD multipole expansion~\cite{Yan:1980uh,Kuang:1981se,Kuang:2006me}. Based on this effective Hamiltonian, the cross section of gluon dissociation in hot medium are derived in the frame of perturbation theory of quantum mechanics~\cite{Liu:2013kkg,Chen:2017jje,Chen:2018dqg}. The result in the Coulomb approximation is in consistent with the OPE .

The goal of this paper is to study the electromagnetic field effect on the gluon dissociation process and the charmonium decay width in QGP. We first introduce in Section~\ref{sec2} the framework of QCD multipole expansion, including an external electromagnetic field. We then systematically solve the two-body Schr\"odinger equation for a pair of charm quarks at finite temperature. At mean field level the solution of the equation gives the magnetic field dependence of the static properties of the $c\bar c$ bound states, shown in Section~\ref{sec3}. Above the mean field we focus in Section~\ref{sec4} on the magnetic field effect on the gluon-dissociation cross-section and calculate the corresponding decay width, by taking the color electric and magnetic dipole interactions as perturbations and employing Fermi's Golden Rule. We summarize in Section~\ref{sec5}.

%==================================
\section{QCD multipole expansion}
\label{sec2}
Multipole expansion is widely used for studying radiation processes in classical electrodynamics~\cite{Bhanot:1979af,Pineda:1997ie,Brambilla:2017uyf}. Considering the large mass and slowly moving of heavy quarks, a heavy flavor system can be treated non-relativistically, and a multipole expansion of the changing gluon field converges rapidly~\cite{Gottfried:1977gp}. The method has been successfully used to calculate hadronic transition rates for both charm and bottom systems~\cite{Gottfried:1977gp,Yan:1980uh,Kuang:1981se}. Including an external electromagnetic field, we start from the gauge-invariant effective Lagrangian density for heavy quarks, which represents the result of partial summation of the perturbation series~\cite{Peskin:1979va,Yan:1980uh} in the absence of electromagnetic field, 
\begin{eqnarray}
\label{Leff}
\mathcal L &=& \int d^3{\bm x}\bar\psi'(x)\left(i\gamma^\mu D_\mu-m_Q\right)\psi'(x)\nonumber\\
&&-{1\over 2}{g^2\over 4\pi}\sum_{a=0}^8\int d^3{\bm x}_1d^3{\bm x}_2\rho_a(x_1){1\over |{\bm r}|}\rho_a(x_2),
\end{eqnarray}
where $m_Q$ is the heavy quark mass, $D_\mu=\partial_\mu+ig\mathcal A_\mu'^a+iqA_\mu$ is the covariant derivative with electric charge $q$, strong coupling constant $g$ and two gauge fields, namely gluon field $\mathcal A_\mu^a$ and photon field $A_\mu$. The interaction among heavy quarks here is perturbatively described by a Coulomb potential between a pair of heavy quarks located at ${\bm x}_1$ and ${\bm x}_2$ with relative coordinate ${\bm r}={\bm x}_1-{\bm x}_2$. To guarantee the gauge invariance, the heavy quark field $\psi(x_i)$ and gluon field $\mathcal A_\mu^a(x_i)$ are transformed to be $\psi'(x_i)=U^{-1}\psi(x_i)$ and $\mathcal A'^a_\mu(x_i)=U^{-1}\mathcal A^a_\mu(x_i) U-(i/g)U^{-1}\partial_\mu U$ through the equal-time gauge link operator $U(x_i)=\mathcal P e^{ig\int_{\bm X}^{{\bm x}_i}d{\bm y}\cdot {\bm {\mathcal A}}^a(y)}$, where $\mathcal P $ is the path-ordering operator and the line integral is along the straight-line segment from the center-of-mass coordinate ${\bm X}=({\bm x}_1+{\bm x}_2)/2$ of the pair to the quark (anti-quark) coordinate ${\bm x}_i$. Note that, the external electromagnetic field $A_\mu(x)$ does not experience such a transformation because it commutates with the link operator $U$. The color charge density (vertex factor) $\rho_a$ is defined as $\rho_a(x_i)=\psi'^\dag(x_i)(\lambda_a/2)\psi'(x_i)$ with Gell-Mann matrix $\lambda_a\ (a=1,...,8$ and $\lambda_0/2=1$) . If the electromagnetic field $A_\mu$ is turned off, the effective Lagrangian becomes the original one in Refs.~\cite{Yan:1980uh,Kuang:1981se}.

The Coulomb potential in the Lagrangian is only the leading term of the color interaction between a pair of heavy quarks. Aiming to go beyond the perturbation theory, one assumes that the heavy quark interaction can be described by a non-relativistic potential and generalizes the Coulomb interaction to including the color confinement (Cornell) part in color singlet state~\cite{Yan:1980uh,Kuang:1981se}. With this consideration, we replace the Coulomb potential $g^2/(4\pi)/|{\bm r}|$ in the above Lagrangian by a general and radial symmetric  potential 
\begin{equation}
\label{potential}
V_a(|{\bm r}|) = V_1(|{\bm r}|)\delta_{a0}+V_2(|{\bm r}|)(1-\delta_{a0}),
\end{equation}
where $V_1$ and $V_2$ are the interaction potentials between $Q$ and $\bar Q$ in color-singlet state and color-octet state.

Using the expression for the gauge link operator $U$, the transformed gluon field can be explicitly expressed as~\cite{Yan:1980uh}
\begin{eqnarray}
\label{gauge}
\mathcal A_0'^a(x_i) &=& \mathcal A_0^a(x_i) + \int_{\bm X}^{{\bm x}_i} d{\bm y} \cdot {\partial {\bm {\mathcal A}}^a(y)\over \partial t},\nonumber\\
{\bm {\mathcal A}}'^a(x_i) &=& {\bm {\mathcal A}}^a(x_i)-{\bm \nabla} \int_{\bm X}^{{\bm x}_i} d{\bm y}\cdot {\bm {\mathcal A}}^a(y),
\end{eqnarray}
and expanding further the original field $\mathcal A_\mu^a$ in Taylor series of ${\bm x}_i-{\bm X}$ at the center-of-mass coordinate ${\bm X}$, one obtains the perturbative expression of $\mathcal A_\mu'^a$ in terms of the color-electric and color-magnetic fields ${\bm {\mathcal E}}^a=\partial {\bm {\mathcal A}}^a/\partial t$ and ${\bm {\mathcal B}}^a={\bm \nabla} \times {\bm {\mathcal A}}^a$, 
\begin{eqnarray}
\label{gauge2}
\mathcal A_0'^a(x_i) &=& \mathcal A_0^a(X) -({\bm x}_i-{\bm X})\cdot {\bm {\mathcal E}}^a(X)+\cdots,\nonumber\\
{\bm {\mathcal A}}'^a(x_i) &=& -({\bm x}_i-{\bm X})\times {\bm {\mathcal B}}^a(X)/2+\cdots.
\end{eqnarray}

The effective Lagrangian (\ref{Leff}) with the non-perturbative interaction (\ref{potential}) is the potential version of QCD to treat heavy quark systems and the foundation for us to calculate the quarkonium gluon-dissociation. If neglecting the color degrees of freedom and the external electromagnetic field, it returns to the QED multipole expansion~\cite{Bhanot:1979af,Pineda:1997ie,Brambilla:2017uyf}. To solve the Schr\"odinger equation for a $Q\bar Q$ system, we transfer the Lagrangian to the Hamiltonian in coordinate representation, 
\begin{eqnarray}
\label{hamiltonian}
\hat H &=& \hat H_0 + \hat H_I,\nonumber\\
\hat H_0 &=& {(\hat {\bm p}_1-q{\bm A}(x_1))^2 \over 2m_Q}+{(\hat {\bm p}_2+q{\bm A}(x_2))^2 \over 2m_Q}\nonumber\\
&&-A_0(x_1)-A_0(x_2)+V_1(|{\bm r}|)+\sum_{a=1}^8{\lambda_a\over 2}{\bar\lambda_a\over 2}V_2(|{\bm r}|), \nonumber\\
\hat H_I &=& q_a \mathcal A^a_0(X)-{\bm d}_a\cdot {\bm {\mathcal E}}^a (X)-{\bm m}_a\cdot {\bm {\mathcal B}}^a(X)+\cdots,
\end{eqnarray}
where $\hat{\bm p}_i=-i{\bm \nabla}_i$ is the heavy quark (anti-quark) momentum operator, and 
\begin{eqnarray}
\label{poles}
q_a &=& g(\lambda_a+\bar\lambda_a)/2,\nonumber\\
{\bm d}_a &=& g({\bm x}_1-{\bm x}_2)(\lambda_a-\bar\lambda_a)/4,\nonumber\\
{\bm m}_a &=& g/m_Q(\lambda_a-\bar\lambda_a)({\bm \sigma}_1-{\bm \sigma}_2)/8
\end{eqnarray}
are the color monopole, electric dipole and magnetic dipole moments of the $Q\bar Q$ system with the Pauli matrix ${\bm \sigma}_i$ for heavy quark and anti-quark. It is clear that, $\hat H_0$ describes a pair of heavy quarks moving in a mean field which contains two parts: the strong potentials $V_1$ and $V_2$ and electromagnetic potential $\mathcal A_\mu$, and $\hat H_I$ is considered as a perturbation above the mean field. The former controls the static properties of the $Q\bar Q$ bound states, and the latter characterizes the quarkonium gluon-dissociation into a color-octet state. 

Focusing on charmonia (bottom quarks are too heavy and probably not so sensitive to the electromagnetic field) and taking the standard perturbative calculation in quantum mechanics, the $c\bar c$ transition rate from a charmonium state into a color octet state via absorbing a gluon at leading order can be given by Fermi’s Golden Rule, $\Gamma=2\pi |_8\langle c\bar c|\hat H_I|\Psi\rangle|^2\rho(E_{c\bar c})$, where $|\Psi\rangle$ and $|c\bar c\rangle _8$ are the initial charmonium bound state and final octet scattering state, and $\rho(E_{c\bar c})$ is the phase-space volume of the final state with energy $E_{c\bar c}$. The transition can be divided into the color-electric dipole ($E1$) and color-magnetic dipole ($M1$) parts. Dividing the transition rate by the flux of the incident gluons, one can obtain the corresponding cross section. Following the procedure in Refs.~\cite{Chen:2017jje,Chen:2018dqg}, the cross sections via transition processes $E1$ and $M1$ read
\begin{eqnarray}
\label{crosssection}
\sigma_{E1} &=& {\pi g^2E_g\over 18} \sum_{n,m,k}|\langle nmk|{\bm r}|\Psi \rangle|^2\delta(E_g-E_B-E_{nmk}),\nonumber\\
\sigma_{M1} &=& {\pi g^2E_g\over 6m_c^2} \sum_{n,m,k}|\langle nmk|\Psi \rangle|^2 \delta(E_g-E_B-E_{nmk})
\end{eqnarray}
with the explicit transition matrix elements
\begin{eqnarray}
\label{element}
\langle nmk|{\bm r}|\Psi\rangle &=& \int d^3{\bm r} \Phi^*_{nmk}({\bm r}){\bm r} \Psi({\bm r}),\nonumber\\
\langle nmk|\Psi\rangle &=& \int d^3{\bm r}\Phi^*_{nmk}({\bm r}) \Psi({\bm r}),
\end{eqnarray}
where $E_g$ is the incident gluon energy, $E_B$ and $\Psi({\bm r})$ are the binding energy and wave function of the charmonium state $|\Psi\rangle$, and $E_{nmk}$ and $\Phi_{nmk}({\bm r})$ are the relative energy and wave function of the $c\bar c$ pair in color-octet state. The $\delta-$function guarantees the energy conservation in the transition processes.

Before we solve the relative motion for the charmonium state and octet state in Section \ref{sec3} and then calculate the charmonium dissociation cross section in Section \ref{sec4}, we simply point out the external electromagnetic field effect on the cross section. While the perturbative Hamiltonian $\hat H_I$ is electromagnetic field independent, the initial and final states $|\Psi\rangle$ and $|c\bar c\rangle _8$ of the transition are both the field dependent. Especially, for the color octet state $|c\bar c\rangle _8$, it is no longer a bound state of strong interaction, but probably a bound state of electromagnetic interaction in the plane perpendicular to the magnetic field~\cite{Chen:2020xsr}. That is the reason why we describe the octet state $|c\bar c\rangle _8=|nmk\rangle$ with two discrete quantum numbers $n$ and $m$ for the transverse bound state and a continuous momentum $k$ for the longitudinal motion. Therefore, the summation over the final state energy means a summation over $n$ and $m$ and an integration over $k$, $\sum_{n,m,k}=\sum_{n,m}\int dk$.      

%==================================
\section{Static properties of $c\bar c$ pairs}
\label{sec3}
Both the charmonium state $|\Psi\rangle$ and octet state $|nmk\rangle$ are determined by the main Hamiltonian $\hat H_0$. We first consider the Schr\"oedinger equation for the charmonium state $|\Psi\rangle$ at finite temperature $T$ and under external magnetic field $B$, 
\begin{equation}
\label{charmonium}
\hat H_0|\Psi\rangle = E|\Psi\rangle.
\end{equation}
Taking the symmetric gauge for electromagnetic field $A_\mu =(-{\bm E}\cdot {\bm x}, ({\bm B}\times{\bm x})/2)$, and making transformation from the coordinates ${\bm x}_1$ and ${\bm x}_2$ to the center-of-mass and relative coordinates ${\bm X}$ and ${\bm r}$ and from the quark momenta ${\bm p}_1$ and ${\bm p}_2$ to their total and relative momenta ${\bm P} = {\bm p}_1 + {\bm p}_2$ and ${\bm p} = ({\bm p}_1 - {\bm p}_2)/2$, the total kinetic energy in $\hat H_0$ becomes
\begin{equation}
\label{kin}
{(\hat{\bm p}_1 - q {\bm A}(x_1))^2\over 2m_c}+{(\hat{\bm p}_2 + q {\bm A}(x_2))^2\over 2m_c} = {\hat{\bm P}_{kin}^2\over 4m_c} + {\hat{\bm p}'^2\over m_c}
\end{equation}
with kinetic momentum ${\bm P}_{kin} = {\bm P}-q{\bm B}\times{\bm r}/2$ and modified relative momentum ${\bm p}' = {\bm p}-q{\bm B}\times{\bm X}/2$. While the kinetic momentum ${\bm P}_{kin}$ and total momentum ${\bm P}$ are not conserved in electromagnetic field with $\left[\hat{\bm P},\hat H_0\right]\neq 0$ and $\left[\hat{\bm P}_{kin},\hat H_0\right]\neq 0$, the pseudo-momentum ${\bm P}_{ps} = {\bm P}+q {\bm B}\times{\bm r}/2$ is a conserved quantity with $\left[\hat{\bm P}_{ps},\hat H_0\right]=0$~\cite{Alford:2013jva}. Keeping this in mind, one factorizes the total wave function as $e^{i({\bm P}_{ps}- q{\bm B}\times{\bm r}/2)\cdot {\bm X}}\Psi({\bm r})$. Substituting this factorization into the Schr\"odinger equation (\ref{charmonium}), one derives the equation controlling the relative energy $E_\Psi=E-{\bm P}_{ps}^2/(4m_c)$ and wave function $\Psi({\bm r})$,
\begin{eqnarray}
\label{relative}
&& \bigg[{\hat {\bm p}^2\over m_c} + {q^2 ({\bm B}\times {\bm r} )^2-2q({\bm P}_{ps}\times {\bm B})\cdot {\bm r}\over 4m_c}\nonumber\\
&& \ \ -{\bm E}\cdot {\bm r}+V_1({\bm r})\bigg]\Psi({\bm r})= E_\Psi\Psi({\bm r}).
\end{eqnarray}
The equation has been solved in previous studies for both charmonium and bottomonium systems~\cite{Marasinghe:2011bt,Alford:2013jva,Machado:2013rta,Cho:2014exa,Guo:2015nsa,Bonati:2015dka,Bonati:2017uvz,Yoshida:2016xgm,Zhao:2020jqu,Mishra:2020kts,Chen:2020xsr,Iwasaki:2021nrz}. 
Considering the fact that, the electromagnetic field breaks down the central symmetry, the orbital angular momentum is no longer conserved even the strong potentials $V_1$ and $V_2$ are radial symmetric. Therefore, one can not further separate the relative wave function into a radial part and the eigen state $Y_{lm}(\theta,\varphi)$ of the orbital angular momentum. In this case a straightforward way to solve the relative equation is to expand the wave function in terms of $Y_{lm}$,
\begin{equation}
\label{expansion}
r\Psi({\bm r})= \sum_{l,m} \phi_{lm}(r) Y_{lm}(\theta,\varphi).
\end{equation}

To simplify the calculation, we consider in the following only magnetic field and neglect the electric field. For convenience, we take the magnetic field to be in the $z$-direction ${\bm B} = B{\bm e}_z$ and the transverse pseudo-momentum in the $y$-direction ${\bm P}_{ps}^\perp = P_{ps}^\perp{\bm e}_y$. Under this choice, the Lorentz potential and the quadratic term in the relative equation become $-q({\bm P}_{ps}\times{\bm B})\cdot{\bm r}/(2m_c)=qBP_{ps}^\perp r\sin\theta\sin\varphi/(2m_c)$ and $q^2({\bm B}\times{\bm r})^2/(4m_c)=q^2B^2r^2\sin^2\theta/(4m_c)$. Expanding the functions $\sin^2\theta Y_{lm}$ and $\sin\theta \sin\varphi Y_{lm}$ in terms of $Y_{lm}$, one obtains the equations for the radial functions $\phi_{lm}(r)$,   
\begin{eqnarray}
\label{radial}
&& \bigg[-{d^2\over d r^2}+m_cV_1(r) + {U\over r^2} +{q^2 B^2V\over 4}r^2 + {qBP_{ps}^\perp W\over 2} r\nonumber\\
&& \ \ -m_cE_\Psi\bigg]R(r) = 0
\end{eqnarray}
with the coefficient matrices
\begin{eqnarray}
\label{coef}
U &=& l(l+1) \delta_{ll'}\delta_{lm'},\nonumber\\
V &=& u_{lm}\delta_{ll'}\delta_{mm'}-v_{lm}\delta_{l+2,l'}\delta_{mm'}-v_{l-2,m}\delta_{l-2,l'}\delta_{mm'},\nonumber\\
W &=& w_{l-1,-m-1}\delta_{l-1,l'}\delta_{m+1,m'}-w_{lm}\delta_{l+1,l'}\delta_{m+1,m'}\nonumber\\
&&+w_{l-1,m-1}\delta_{l-1,l'}\delta_{m-1,m'}-w_{l,-m}\delta_{l+1,l'}\delta_{m-1,m'},\nonumber\\
u_{lm} &=& {2(l^2+l-1+m^2)\over (2l-1)(2l+3)},\nonumber\\
v_{lm} &=& {1\over 2l+3}{\sqrt{((l+1)^2-m^2)((l+2)^2-m^2)}\over \sqrt{(2l+1)(2l+5)}},\nonumber\\
w_{lm} &=& {\sqrt{(l+m+1)(l+m+2)}\over 2i\sqrt{(2l+1)(2l+3)}}
\end{eqnarray} 
and the radial wave function vector
\begin{equation}
\label{matrix}
R(r) = (\phi_{00}(r),\phi_{1,-1}(r),\phi_{10}(r),\phi_{11}(r),...)^T.
\end{equation}
Since the matrices $V$ and $W$ are with off-diagonal elements, this is a group of coupled equations for, in principle, all the radial functions. In a realistic calculation, a cut-off of the orbital angular momentum is needed, $l \leq l_{max}$. We choose $l_{max}=7$ and solve the radial equation via the inverse power method~\cite{H.W. Crater}. 

\begin{figure}[!htb]
\includegraphics[width=0.39\textwidth]{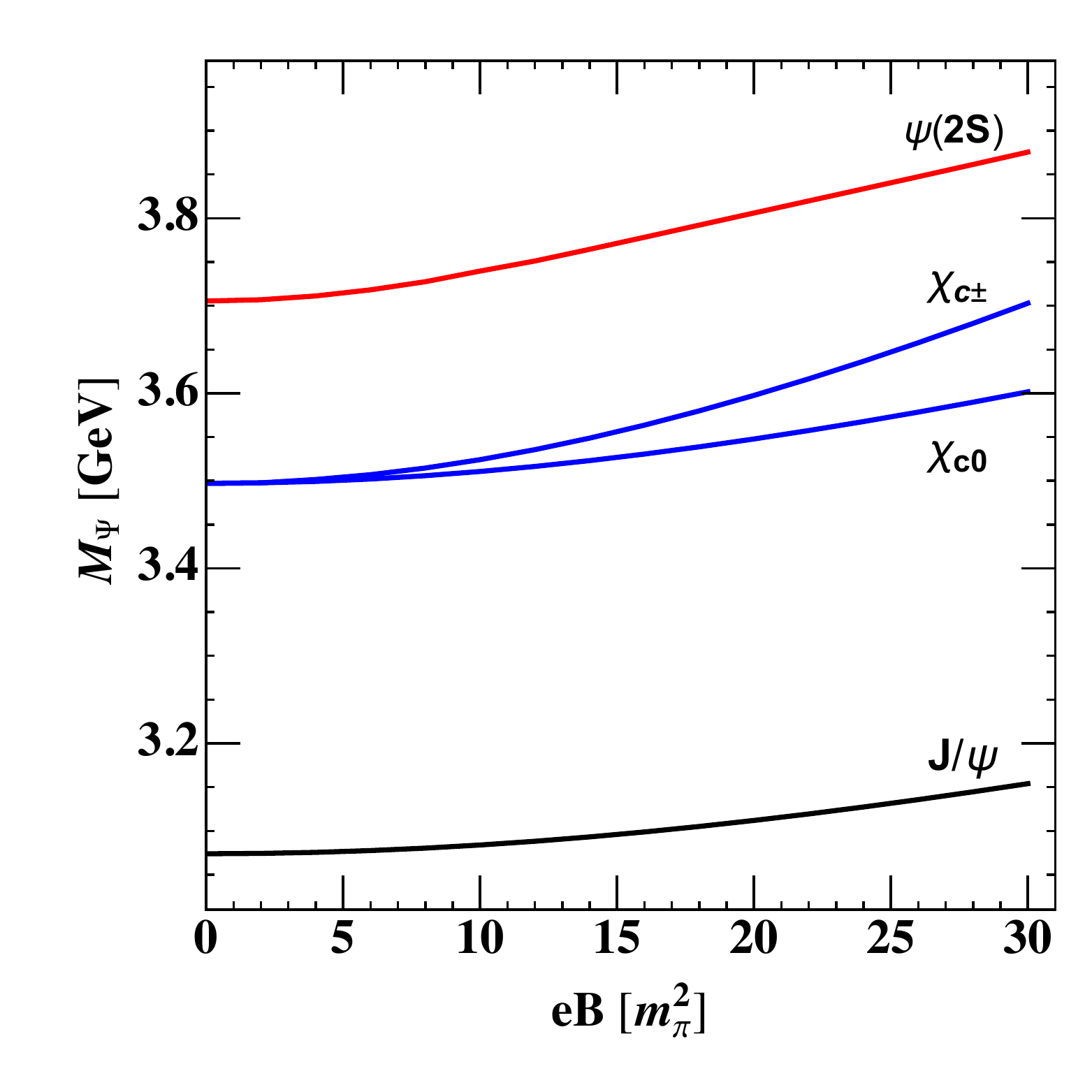}
\caption{The charmonium mass $M_\Psi$ as a function of magnetic field $eB$ at vanishing temperature and pseudo-momentum $T=0$ and ${\bm P}_{ps}=0$. }
\label{fig1}
\end{figure}

Like usual treatment~\cite{Liu:2013kkg,Chen:2017jje,Chen:2018dqg}, we have neglected in the relative equation the potential $V_2$ in color octet state. In vacuum the potential $V_1$ in color singlet state is often taken as the Cornell form,
\begin{equation}
\label{cornell}
V_1(|{\bm r}|)=-{\alpha\over |{\bm r}|}+\sigma |{\bm r}|.
\end{equation}
The eigen value of the radial equation (\ref{radial}) determines the charmonium mass $M_\Psi=2m_c+E_\Psi$ at zero temperature. Taking the charm quark mass $m_c=1.29$ GeV, by fitting the experimentally measured charmonium masses at vanishing magnetic field, the two parameters in the potential are fixed to be $\alpha=0.4105$ and $\sigma=0.2~\text{GeV}^2$~\cite{Zhao:2020jqu}. When the magnetic field is turned on, the central symmetry is broken by the field, the energy levels of the P-wave states with different magnetic quantum number $m$ will no longer degenerate. For instance, the $\chi_c$ state split into three states $\chi_{c0}$, $\chi_{c+}$, and $\chi_{c-}$, corresponding to the magnetic quantum number $m=0,1,-1$. On the other hand, if we take the conserved pseudo-momentum ${\bm P}_{ps}=0$, the rotational symmetry around the $z$-axis is restored, which leads to the degeneration of the two states $\chi_{c+}$ and $\chi_{c-}$. The masses of $J/\psi$, $\psi(2S)$, $\chi_{c0}$, and $\chi_{c\pm}$ states are shown in Fig.~\ref{fig1}. It is clear that all the charmonium masses increase with the magnetic field, due to the attractive quadratic potential in the relative equation (at ${\bm P}_{ps}=0$ this is the only electromagnetic potential). The result is similar to the previous study~\cite{Alford:2013jva}.
%---------------------------------------------------------------------
%---------------------------------------------------------------------

We now turn to the calculation at finite temperature. Due to the many-body interaction in hot medium, the potential between $c$ and $\bar c$ is screened. When the screening length (screening mass) is short (large) enough, the charmonium state is melted by the medium. At very high temperature, the hard-thermal loop (HTL) calculation shows that the potential is modified by a screening factor $e^{-m_D r}$ with the Debye mass $m_D$~\cite{Laine:2006ns}. For the QGP at finite temperature the potential is simulated by lattice QCD~\cite{Burnier:2014ssa,Burnier:2015tda}. Based on the Gauss-law approach by using the permittivity obtained from the HTL approximation to modify the non-perturbative vacuum potential, one takes the finite temperature potential $V_1$ as~\cite{Lafferty:2019jpr}, 
\begin{eqnarray}
V_1(T,r) &=& -\alpha\left[m_D+{e^{-m_Dr}\over r}\right]\nonumber\\
&& +{\sigma\over m_D}\left[2-(2+m_Dr)e^{-m_Dr}\right],
\end{eqnarray}
and the temperature dependent Debye mass $m_D(T)$ is obtained by fitting the lattice data~\cite{Burnier:2014ssa,Burnier:2015tda}. The influence of the magnetic field on Debye mass is neglected here, since the change is very small~\cite{Singh:2017nfa,Hasan:2018kvx,Hasan:2017fmf}.

%---------------------------------------------------------------------
\begin{figure}[!htb]
\includegraphics[width=0.39\textwidth]{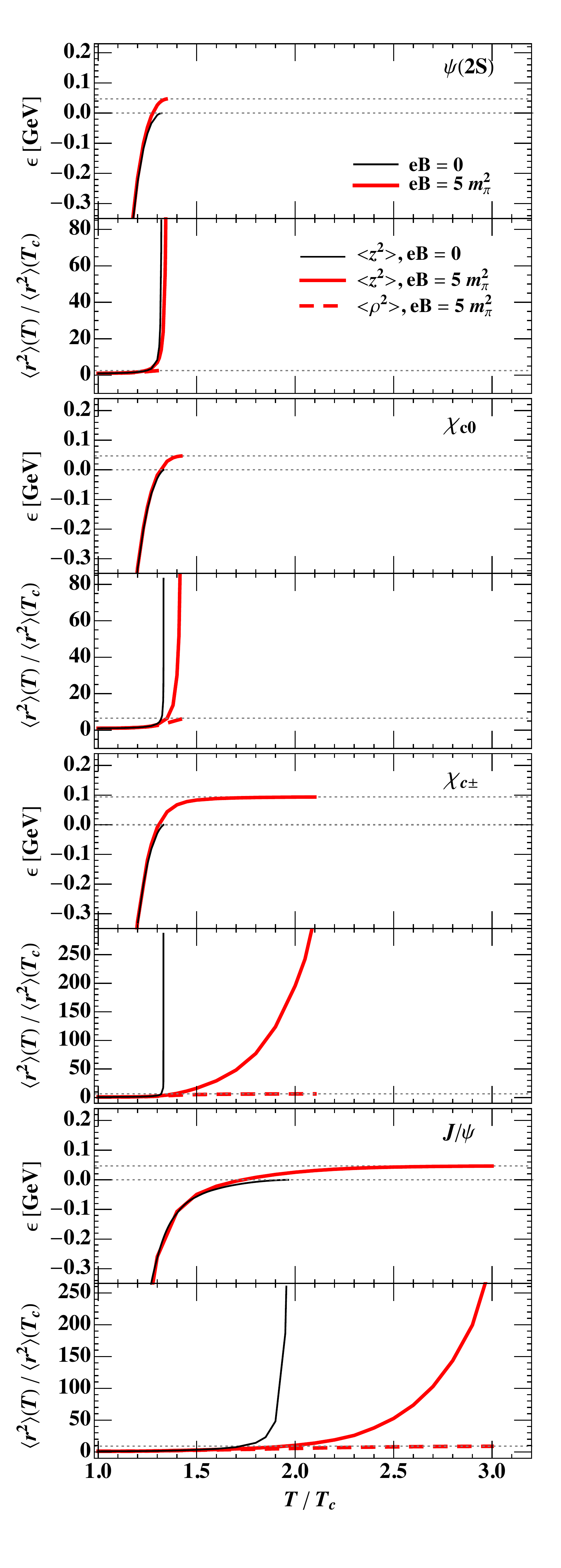}
\caption{The charmonium binding energy $\epsilon$ and longitudinal and transverse mean square radii $\langle z^2\rangle$ and $\langle \rho^2\rangle$ as functions of temperature $T$ at vanishing pseudo-momentum ${\bm P}_{ps}=0$. The temperature and radii are scaled by their values at the deconfinement phase transition temperature $T_c$. The thin and thick solid lines are $\epsilon$ and $\langle z^2\rangle$ at $eB=0$ and $5~m_\pi^2$, and the dashed line is $\langle \rho^2\rangle$ at $eB=5~m_\pi^2$. }
\label{fig2}
\end{figure}
%---------------------------------------------------------------------
At finite temperature, the long-distance part of the potential is suppressed by the hot medium and becomes saturated with the value $V_1(T,\infty)=-\alpha m_D+2\sigma/m_D$. Therefore, the charmonium binding energy relative to the saturated potential is redefined as $\epsilon=E_\Psi-V_1(T,\infty)$. The temperature and magnetic field dependence of the binding energy and mean square radii $\langle z^2\rangle$ in longitudinal direction and $\langle \rho^2\rangle=\langle x^2\rangle+\langle y^2\rangle$ in transverse plane are shown in Fig.~\ref{fig2}, again the conserved pseudo-momentum is taken to be zero ${\bm P}_{ps}=0$. Since what we are interested in is the charmonium behavior in the QGP phase, the temperature we considered here is above the critical temperature $T_c=172$ MeV~\cite{Lafferty:2019jpr} of deconfinement phase transition. Let's first consider the pure temperature effect, see the thin solid lines. The binding energy, which is negative, approaches to zero gradually and becomes saturated at the melting temperature $T_m$ with $\epsilon(T_m)=0$. Correspondingly, the mean square radii $\langle z^2\rangle$ and $\langle \rho^2\rangle$, which are the same due to the radial symmetry of the system in the absence of magnetic field, increase with temperature and go to infinity at $T_m$. Obviously, the excited states $\psi(2S)$ and $\chi_c$ are easier to be melted than the ground state $J/\psi$, and the three $P-$wave states $\chi_{c0}$ and $\chi_{c\pm}$ are degenerate in the absence of magnetic field.  

Different from the strong interaction ($V_1$) which is suppressed by the hot medium, the external magnetic field is temperature independent, and its effect on the $c\bar c$ pair above the melting temperature $T_m$ becomes the dominant interaction. When the magnetic field is turned on, while the mean square radius $\langle z^2\rangle$ still goes to infinity at high enough temperature, the magnetic interaction confines the pair motion in the transverse plane and makes the mean square radius $\langle \rho^2\rangle$ finite at any temperature~\cite{Chen:2020xsr}. Therefore, the melting temperature $T_m$ is in fact a transition temperature for the $c\bar c$ pair to change from a bound state of strong interaction to a transverse bound state of electromagnetic interaction. The melting temperature (transition temperature) $T_m$ can then be defined through the divergence of the longitudinal size $\langle z^2\rangle (T_m)\to \infty$ and the saturation of binding energy and transverse size $\epsilon(T\ge T_m)=const$ and $\langle \rho^2\rangle(T\ge T_m)=const$, see the horizontal lines in Fig.~\ref{fig2}. 

To determine the saturation values, we now turn to calculate the relative energy and wave function $E_{nmk}$ and $\Phi_{nmk}$ for the octet state of $c\bar c$ pairs. When the strong interaction potential $V_1$ disappears, the $c\bar c$ pair is controlled only by the magnetic field. The relative Hamiltonian can be written as 
\begin{equation}
\label{hamil}
{p_x^2\over m_c}+{q^2 B^2\over 4m_c}x^2 + {p_y^2\over m_c}+{q^2 B^2\over 4m_c}\left(y-{P_{ps}^\perp \over qB}\right)^2 + {p_z^2\over m_c}.
\end{equation}
It is clear that the relative motion can be separated into a two dimensional harmonic oscillator in the $x-y$ plane and a plain wave in the $z$ direction. The eigen value $E_{nmk}$ and eigen function $\Phi_{nmk}({\bm r})$ of the Hamiltonian can analytically be expressed as
\begin{eqnarray}
\label{eigen}
&& E_{nmk} = (2n+|m|+1){qB\over m_c} + {k^2\over m_c}, \\
&& \Phi_{nmk}({\bm r}) = N_{nm} {e^{ikz}\over \sqrt{2\pi}}\rho^{|m|} L_n^{(|m|)}(qB\rho^2/2)e^{-qB\rho^2/4}e^{im\varphi}\nonumber
\end{eqnarray}
with the normalization factor $N_{nm} = \sqrt{n! (qB/2)^{|m|+1}/(n+|m|)!/\pi}$, where $k$ is the continuous momentum describing the plane wave in the $z-$direction, the main and magnetic quantum numbers $n$ and $m$ characterize the transverse wave function, and $L_n^{(|m|)}$ is the associated Laguerre polynomials. The transverse radius $\rho$ and azimuth angle $\varphi$ are defined through $x = \rho \cos\varphi$ and $y = P_{ps}^\perp/(qB) + \rho \sin\varphi$, and the wave function satisfies the orthogonal condition
\begin{equation}
\int d^3{\bm r}\Phi^*_{nmk}({\bm r}) \Phi_{n'm'k'}({\bm r}) = \delta_{nn'}\delta_{mm'} \delta(k-k').
\end{equation}

With the relative energy level $E_{nmk}$, one can determine the saturation values of the binding energy $\epsilon$ and transverse mean squared radius $\langle\rho^2\rangle$ of the charmonium state. They are controlled by the corresponding lowest Landau energy level, 
\begin{eqnarray}
\label{saturation}
&&\epsilon(T_m) =(1+|m|) {qB \over m_c}, \nonumber\\
&&\langle\rho^2\rangle(T_m) = (1+|m|){2\over qB}.
\end{eqnarray}
The saturated binding energy increases linearly with the magnetic field, and self-consistently the saturated transverse size decreases linearly with the field, which mean a more and more tight $c\bar c$ bound state of electromagnetic interaction in the transverse plane.    

%%%%%%%%%%%%%%%%%%%%%%%%%%%%%%%%%%%%%%
\section{Charmonium gluon-dissociation}
\label{sec4}
To calculate the gluon dissociation cross sections (\ref{crosssection}), we need the wave functions $\Psi$ and $\Phi_{nmk}$ and the binding energies $E_B$ and $E_{nmk}$ for the initial charmonium and final octet states. $\Psi, \Phi_{nmk}$ and $E_{nmk}$ are calculated in the last section. The charmonium binding energy at finite temperature $\epsilon(T)=E_\Psi(T)-V_1(T,\infty)$ is relative to the saturated strong potential. Considering the fact that the electromagnetic interaction makes the binding energy nonzero above the dissociation temperature, the charmonium binding energy $E_B$ defined through the energy conservation in dissociation cross sections (\ref{crosssection}) should be  
\begin{eqnarray}
\label{eb}
E_B(T) &=& -\left[\epsilon(T)-\epsilon(T_m)\right]\nonumber\\
&=& -\left[E_\Psi(T)-V_1(T,\infty)-\epsilon(T_m)\right],
\end{eqnarray}
when both the strong and electromagnetic interactions are taken into account. In this case the binding energy $E_B$ satisfies the physics: it vanishes above the dissociation temperature, $E_B(T>T_m)=0$.  

When the charmonia are at rest with ${\bm P}_{ps} = 0$, the expansion for the relative wave function (\ref{expansion}) is reduced to  
\begin{equation}
\label{swave}
\Psi({\bm r}) = \sum_{l=0}^\infty \sqrt{{2l+1\over 4\pi}} \phi_l(r) P_l(\cos \theta)
\end{equation}
for the $S$-wave states $J/\psi$ and $\psi(2S)$ with even $l$ and $P-$wave state $\chi_{c0}$ with odd $l$, and 
\begin{equation}
\label{pwave}
\Psi({\bm r}) = \sum_{l=1}^\infty \sqrt{{2l+1\over 4\pi l(l+1)}} \phi_l(r) P_l^{(1)}(\cos \theta) (e^{-i \varphi} \pm e^{i \varphi} )
\end{equation}
for the $P-$wave states $\chi_{c\pm}$ with odd $l$.
	
Substituting the expansion (\ref{swave}) for $J/\psi$, $\psi(2S)$ and $\chi_{c0}$ into the transition elements (\ref{element}) and using the explicit expression for the octet state $\Phi_{nmk}({\bm r})$ (\ref{eigen}) with $\rho=r\sin\theta$ and $z=r\cos\theta$, the integration over the azimuth angle $\varphi$ leads to the selection rules: the transition elements $\langle nmk|z|\Psi\rangle$ and $\langle nmk|\Psi\rangle$ are always zero unless $m=0$, and the elements $\langle nmk|x|\Psi\rangle$ and $\langle nmk|y|\Psi\rangle$ are always zero unless $m=\pm 1$. Since gluon carries spin $1$ and its $z$ component is $1$, $0$ and $-1$, the physics behind the selection rules is the conservation of the $z$ component of total angular momentum for charmonium states with zero $z$ component of orbital angular momentum. From the $m-$dependence of the wave function $\Phi_{nmk}$, the nonzero transition elements depend only on $|m|$. This means that there are only one independent transition element $T_n$ for channel $M1$ and two independent elements $T_{nz}$ and $T_{n\rho}$ for channel $E1$, 
\begin{eqnarray}
\label{three}
&& T_n(k) = N_{n0} \sum_{l,r,x} r^2 G_{nl}^{(0)}(r,x,k),\\
&& T_{nz}(k) = N_{n0} \sum_{l,r,x} r^3 x G_{nl}^{(0)}(r,x,k),\nonumber\\
&& T_{n\rho}(k) = N_{n1} \sum_{l,r,x} r^4 (1-x^2)G_{nl}^{(1)}(r,x,k)\nonumber
\end{eqnarray}
with the definition of $\sum_{l,r,x}=\sum_l\int_0^\infty dr\int_{-1}^1dx$ and 
\begin{eqnarray}
\label{g}
G_{nl}^{(i)}(r,x,k) &=& \sqrt{2l+1\over2} \phi_l(r) e^{-qBr^2(1-x^2)/4}\nonumber\\
&&\times L_n^{(i)}(qBr^2(1-x^2)/2) P_l(x) e^{ikrx}
\end{eqnarray}
for $i=0,1$.

We take then the integration over the longitudinal momentum $k$ in the dissociation cross sections. By employing the relation for the $\delta-$function,
\begin{eqnarray}
\label{delta}
&& \int dk F(k)\delta(E_g-E_B-E_{nmk})\nonumber\\
&=& {m_c\over 2k_{nm}}\left[F(k_{nm}) + F(-k_{nm})\right]
\end{eqnarray}
with
\begin{equation}
k_{nm} = \sqrt{m_c(E_g-E_B-E_{nm0})}
\end{equation}
for any function $F(k)$, the cross sections (\ref{crosssection}) for charmonium states $J/\psi$, $\psi(2S)$ and $\chi_{c0}$ are simplified as
\begin{eqnarray}
\label{crosssection2}
\sigma_{E1} &=& {\pi g^2 E_g\over 18} \sum_n \left[{m_c\over k_{n0}}|T_{nz}(k_{n0})|^2+{m_c\over k_{n1}} |T_{n\rho}(k_{n1})|^2 \right],\nonumber\\
\sigma_{M1} &=& {\pi g^2 E_g\over 6m_c^2} \sum_n {m_c\over k_{n0}}	|T_n(k_{n0})|^2.
\end{eqnarray}
	
For the $P-$wave states $\chi_{c\pm}$, similar calculations can be done. Substituting the expansion (\ref{pwave}) into the transition elements (\ref{element}), the integration over the azimuth angle $\varphi$ is controlled by the selection rules: only for the quantum number $m=\pm 1$ the transition elements $\langle nmk|z|\Psi\rangle$ and $\langle nmk|\Psi\rangle$ are not zero, and only for $m=0$ and $\pm 2$ the elements $\langle nmk|x|\Psi\rangle$ and $\langle nmk|y|\Psi\rangle$ are not zero. The physics is again the conservation of the $z$ component of total angular momentum for charmonium states with $z$ component of orbital angular momentum $\pm 1$. Again the nonzero transition elements are $|m|-$dependent, there are only one independent transition element $\mathcal T_n$ for channel $M1$ and three independent elements $\mathcal T_{nz}, \mathcal T_{n\rho}^{(0)}$ and $\mathcal T_{n\rho}^{(2)}$ for channel $E1$, 
\begin{eqnarray}
\label{four}
&& \mathcal T_n(k) = N_{n1} \sum_{l,r,x} r^3 (1-x^2)^{1/2}\mathcal G_{nl}^{(1)}(r,x,k),\nonumber\\
&& \mathcal T_{nz}(k) = N_{n1} \sum_{l,r,x} r^4 (1-x^2)^{1/2}\mathcal G_{nl}^{(1)}(r,x,k),\nonumber\\
&& \mathcal T_{n\rho}^{(0)}(k) = N_{n0} \sum_{l,r,x} r^3 (1-x^2)^{1/2}\mathcal G_{nl}^{(0)}(r,x,k),\nonumber\\
&& \mathcal T_{n\rho}^{(2)}(k) = N_{n2} \sum_{l,r,x} r^5 (1-x^2)^{3/2}\mathcal G_{nl}^{(2)}(r,x,k)
\end{eqnarray}
with
\begin{eqnarray}
\label{g2}
\mathcal G_{nl}^{(i)}(r,x,k) &=& \sqrt{2l+1\over 2l(l+1)} \phi_l(r) e^{-qBr^2(1-x^2)/4}\\
&&\times L_n^{(i)}(qBr^2(1-x^2)/2) P_l^{(1)}(x) e^{ikrx}\nonumber
\end{eqnarray}
for $i=0,1,2$.

After the integration over the longitudinal momentum $k$, the dissociation cross sections for charmonium states $\chi_{c\pm}$ are expressed as
\begin{eqnarray}
\label{crosssection3}
\sigma_{E1} &=& {\pi g^2 E_g\over 18} \sum_n \bigg[{2m_c\over k_{n1}}|\mathcal T_{nz}(k_{n1})|^2+{m_c\over k_{n2}} |\mathcal T_{n\rho}^{(2)}(k_{n2})|^2\nonumber\\
&& +{m_c\over k_{n0}} |\mathcal T_{n\rho}^{(0)}(k_{n0})|^2\bigg],\nonumber\\
\sigma_{M1} &=& {\pi g^2 E_g\over 6m_c^2} \sum_n {2m_c\over k_{n1}}	|\mathcal T_n(k_{n1})|^2.
\end{eqnarray}
	
%---------------------------------------------------------------------
\begin{table}
	\renewcommand\arraystretch{1.8}
	\setlength{\tabcolsep}{1.0mm}
	\begin{tabular}{c|c|c|c}
		\toprule[1pt]\toprule[1pt] 
		\multicolumn{1}{c|}{} & \multicolumn{1}{c|}{$J/\psi$, $\psi(2S)$} &   \multicolumn{1}{c|}{$\chi_{c0}$} &   \multicolumn{1}{c}{$\chi_{c\pm}$} \tabularnewline
		\midrule[1pt]
		$\sigma_{E1}\propto$ & $1/k_{n1}$ & $1/k_{n0}$ & $1/k_{n0},\ 1/k_{n2}$ \tabularnewline
		$\sigma_{M1}\propto$ & $1/k_{n0}$ & $k_{n0}$ & $1/k_{n1}$ \tabularnewline
		\bottomrule[1pt]\bottomrule[1pt]
	\end{tabular}
	\caption{The charmonium gluon-dissociation cross-sections in channels $E1$ and $M1$ around the maximum Landau energy level. }
	\label{table1}
\end{table}
%--------------------------------------------------------------------
We now analyze the infrared divergence of the transition elements $T$ and $\mathcal T$ in the limit of longitudinal momentum $k_{nm}=0$. Let's consider the $S-$wave states $J/\psi$ and $\psi(2S)$ as an example. In this case, $l$ is even, $P_l(x)$ is an even function, the requirement that the integrated function in any $T$ should be an even function of $x$ leads to the replacement of $e^{ik_{nm}rx}$ by $\cos(k_{nm}rx)$ in $T_n$ and $T_{n\rho}$ and by $i\sin(k_{nm}rx)$ in $T_{nz}$. Around $k_{nm}=0$, by taking the expansions $\cos(k_{nm}rx)=1+\mathcal{O}(k_{nm}^2)$ and $\sin(k_{nm}rx)=k_{nm}rx+\mathcal{O}(k_{nm}^3)$, $\sigma_{E1}$ is proportional to $1/k_{n1}$ and becomes divergent at $k_{n1}=0$, and $\sigma_{M1}$ to $1/k_{n0}$ and divergent at $k_{n0}=0$. Now the only thing left is the condition for the limit $k_{nm}=0$. For a given incident gluon energy $E_g$, the limit is realized only when the maximum Landau energy level $E_{n_{max}m0}=(2n_{max}+|m|+1)qB/m_c$ satisfies the energy conservation,
\begin{equation}
\label{cons}
E_g-E_B-E_{n_{max}m0}=0.
\end{equation}   
The conclusion is therefore the following: When the maximum Landau energy level $E_{n_{max}m0}(E_g)$ satisfies the conservation law, the cross section is divergent at the corresponding $E_g$; If not, the cross section is finite but still peaks at $E_g$. The similar analysis can be done for the $P-$wave states $\chi_{c0}$ and $\chi_{c\pm}$. The behavior of the cross sections around the maximum Landau energy level for all the charmonium states is shown in Table \ref{table1}. Except for channel $M1$ for $\chi_0$, all the other cross sections are divergent at $k_{n0}=0$ or $k_{n1}=0$ or $k_{n2}=0$.
%---------------------------------------------------------------------
\begin{figure}[!htb]
	\includegraphics[width=0.39\textwidth]{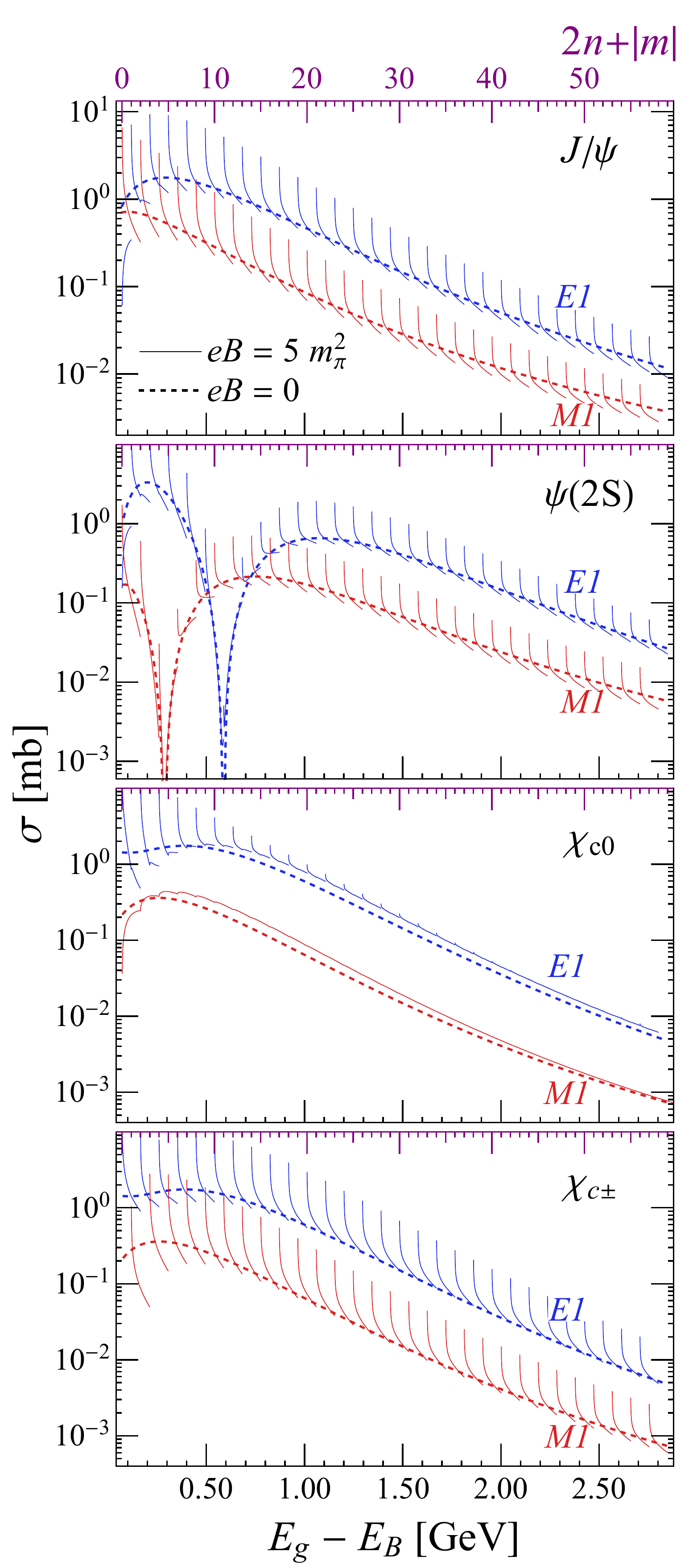} 
	\caption{The charmonium gluon-dissociation cross-sections in channels $E1$ and $M1$ at vanishing temperature and conserved momentum $T=0$ and ${\bm P}_{ps}=0$. The bottom label $E_g-E_B$ is the energy difference between the initial gluon and charmonium, and the top label is $2n+|m|$ characterizing the Landau energy level. Solid and dashed lines are the calculations with and without magnetic field. }
	\label{fig3}
\end{figure}
%---------------------------------------------------------------------

The cross sections in channels $E1$ and $M1$ for different charmonium states at vanishing temperature and conserved momentum are shown in Fig.~\ref{fig3} as functions of incident gluon energy $E_g$. The dashed lines are the result without magnetic field, which were calculated in Ref.~\cite{Chen:2017jje}. When the magnetic field is turned on, while the global trend of the cross section is similar to the one without the field, a significant change is the field induced hair structure. Let us consider $J/\psi$ as an example. As analyzed above, the cross section $\sigma_{E1}\ (\sigma_{M1})$ goes to infinity when the energy difference $E_g-E_B$ between the initial gluon and $J/\psi$ reaches some Landau energy level $2(n+1)qB/m_c\ ((2n+1)qB/m_c)$ characterized by the main quantum number $n$. Therefore, with increasing gluon energy, the cross sections become divergent at the Landau levels and are continuous between two neighboured levels. This indicates that the magnetic field makes the cross sections grow hair!  The behavior of the cross sections for $\psi(2S)$ and $\chi_{c\pm}$ are very similar to $J/\psi$. The only exception is $\chi_{c0}$. As shown in Table \ref{table1}, there is no infrared divergence for the cross section in channel $M1$, $\sigma_{M1}$ is continuous at any incident gluon energy. Note that, the cross sections for the three $P-$wave states $\chi_{c0}$ and $\chi_{c\pm}$ are the same in the absence of magnetic field but separated by the field. 

We finally calculate the charmonium decay width through gluon dissociation at finite temperature and magnetic field. For a charmonium at rest in hot medium, the width is the integration of the weighted cross-section over the gluon momentum,    
\begin{equation}
\label{width}
\Gamma(T,B)=d_g\int {d^3{\bm p}\over (2\pi)^3}\sigma(E_g,T,B)f_g(E_g,T),
\end{equation}
where $d_g,\ {\bm p},\ E_g$ and $f_g$ are the gluon degeneracy, momentum, energy and phase-space distribution. Gluons are massless in vacuum with energy $E_g=|{\bm p}|$ but obtain thermal mass at finite temperature $m_g(T)=\sqrt{(2N_c+N_f)/12}\,gT$~\cite{FTFT} with energy $E_g=\sqrt{{\bm p}^2+m_g^2}$. We take in the calculation the degeneracy $d_g=16$ and coupling constant $g\approx 2$ for $N_c=N_f=3$, as used in Ref.~\cite{Riek:2010fk}. Since gluons do not carry electric charge, the mass and in turn the energy and distribution function are magnetic field independent at leading order (in general the field can change the gluon properties through modifications from quark loops). Therefore, the gluon distribution can be taken as the Bose--Einstein function in the local rest frame of the medium $f_g(E_g,T)=1/(e^{E_g/T}-1)$.  

\begin{figure}[!htb]
	\includegraphics[width=0.39\textwidth]{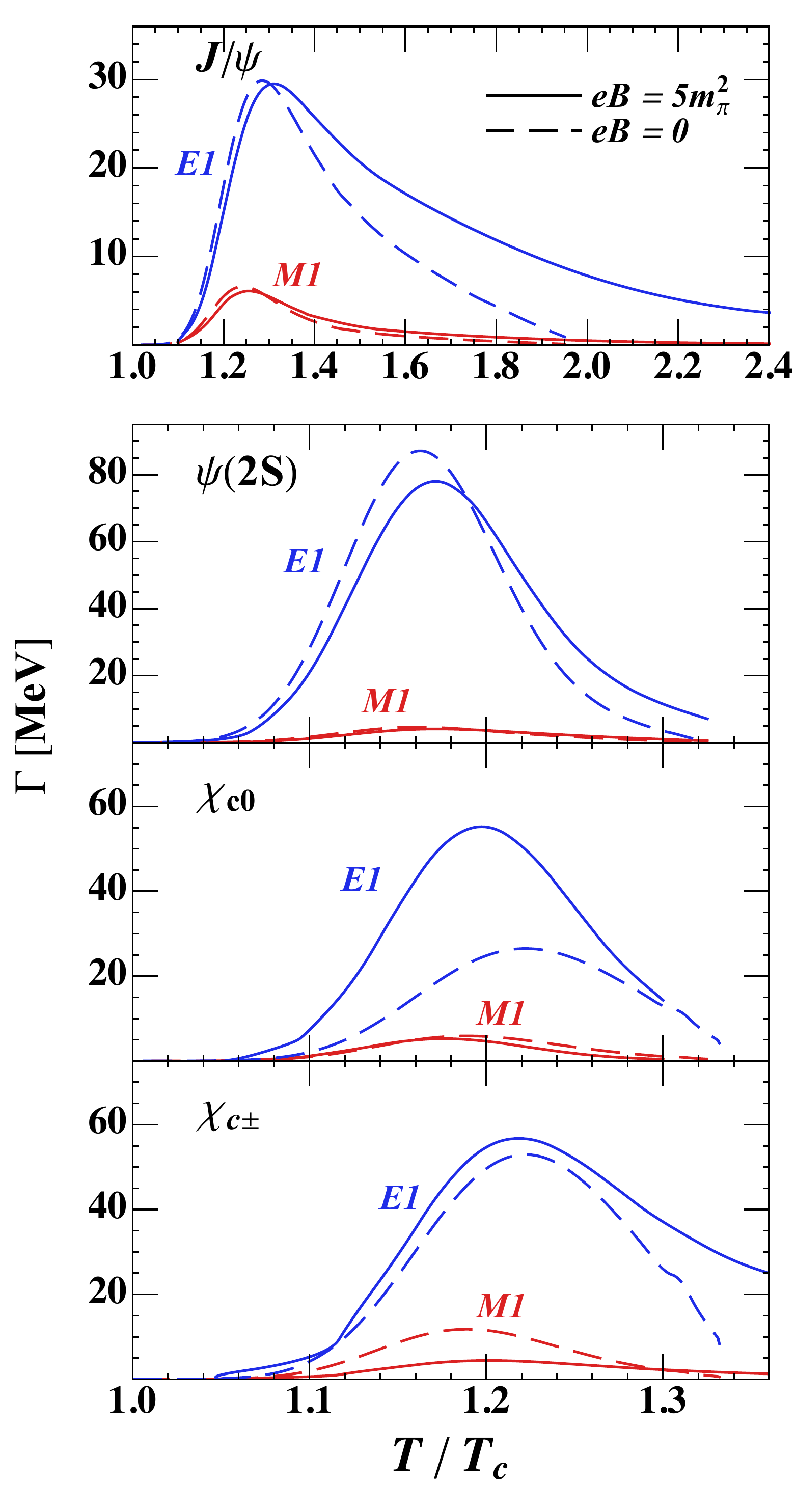}
	\caption{The charmonium decay width through gluon dissociation in channels $E1$ and $M1$ as functions of scaled temperature $T/T_c$ at vanishing pseudo-momentum ${\bm P}_{ps}=0$. Solid and dashed lines are the calculations with and without magnetic field.}
	\label{fig4}
\end{figure}

The charmonium decay widths for channels $E1$ and $M1$ are shown in Fig.~\ref{fig4} as functions of temperature in the deconfined phase with $T>T_c$. From the picture of color screening, the shape of a width is exactly a $\delta-$function located at the melting temperature $T_m$. Considering realistic collision processes, the $\delta-$function is expanded to be a distribution covering both $T<T_m$ and $T>T_m$. While $T_m$ is very different for the ground and excited states, for instance at $eB=5~m_\pi^2$ there are from Fig.\ref{fig2} $T_m/T_c\sim 1.4$ for $\psi(2S)$ and $\chi_{c0}$, $2.1$ for $\chi_{c\pm}$ and $3$ for $J/\psi$, all the decay widths peak at about $T/T_c\sim 1.2$. For any charmonium state and in any case with and without magnetic field, the channel $E1$ always dominants both the cross section and the decay width, in comparison with the channel $M1$. This is mainly due to the $M1$ suppression by the mass factor $m_c^2$ in the denominator of the cross sections, see Eqs.~(\ref{crosssection2}) and (\ref{crosssection3}). It is also easy to understand that the loosely bound states $\psi(2S)$ and $\chi_c$ are easier to decay than the tightly bound state $J/\psi$.     

%---------------------------------------------------------------------
%---------------------------------------------------------------------
Now we focus on the magnetic field effect on the decay width. Considering the fact that $k$ is the magnitude of the charmonium longitudinal momentum, $k_{nm}=\sqrt{m_c(E_g(p)-E_B-E_{nm0})}$ should be positive, and the momentum integration (\ref{width}) around a divergence is proportional to 
\begin{equation}
\label{int}
\int dp{1\over k_{nm}}\theta(k_{nm}) = 2\int_0^\delta dk_{nm}=2\delta
\end{equation}
and is finite. Therefore, the integrated decay width is convergent at any temperature $T$. Secondly, the radial symmetry breaking deforms the charmonium and octet states, the change in the transition element $\langle nmk|{\bm r}|\Psi\rangle$ by the magnetic field should be stronger than the element $\langle nmk|\Psi\rangle$. This means that the cross section and decay width in channel $E1$ are more sensitive to the field than that in channel $M1$. Due to the larger deformation of the $P-$wave states, the magnetic field effect on $\chi_{c0}$ and $\chi_{c\pm}$ is more important than the $S-$wave states $J/\psi$ and $\psi(2S)$. This is clearly shown in Fig.~\ref{fig4}.     

%==============
\section{Summary}
\label{sec5}
A typical quantum mechanics problem is the particle motion in a magnetic field, which leads to the famous Landau energy levels. While the magnetic field effect is recently widely discussed in high energy physics, like the influence on QCD phase transitions and static particle properties, it is rarely introduced in the calculation of particle collisions. In this paper we investigated the gluon dissociation process $g+\Psi\to c+\bar c$ in a strong magnetic field and found that the Landau energy levels make the cross section grow hair.

We extended the QCD multipole expansion for a pair of heavy quarks to including an external electromagnetic field. By solving the two-body Schr\"odinger equation with mean field potentials for strong and electromagnetic interactions we firstly determined the charmonium static properties, including the binding energy and wave function. Taking then the color dipole interactions as perturbations above the mean field and employing Fermi's Golden Rule, we focused on the magnetic field effect on the gluon-dissociation process in the quark-gluon plasma. In general case the dissociation cross-section becomes divergent when the energy difference between the initial gluon and charmonium reaches a Landau energy level for the final octet state. These divergences at different Landau levels look like hairs of the cross section. However, the gluon energy integrated decay width is always continuous at any temperature. Considering the deformation of the charmonium states, especially for the loosely bound states, the magnetic field effect on the color-electric dipole channel and the excited states is significantly important. In our numerical calculation the difference between the decay widths with and without magnetic field is already large enough at $eB=5~m_\pi^2$. This indicates that the magnetic field effect on charmonium dissociation in high energy nuclear collisions at RHIC and LHC energies might be sizeable and considered as a probe of the initially produced electromagnetic field.        

{\bf Acknowledgement:} We thank very much Shile Chen and Kai Zhou for helpful discussions. The work is supported by NSFC grant Nos. 11890712, 12035006, 12047535 and 12075129 and Guangdong Major Project of Basic and Applied Basic Research No. 2020B0301030008. S.S. is grateful to supports from Natural Sciences and Engineering Research Council of Canada, the Bourses d'excellence pour \'etudiants \'etrangers (PBEEE) from Le Fonds de Recherche du Qu\'ebec - Nature et technologies (FRQNT), and the U.S. Department of Energy, Office of Science, Office of Nuclear Physics under grant No. DEFG88ER40388.

%--------------------------------------------------------------------------------------------------------------------------------

\end{document}